\documentclass[psfig,referee]{aa} 

 \usepackage{graphics}
\begin{document}

\title{The velocity structure of moving magnetic feature pairs
around sunspots: support for the U-loop model}

\author{Jun Zhang\inst{1, 2}, S. K. Solanki\inst{1}, J. Woch\inst{1},
\and Jingxiu Wang\inst{2}}

\offprints{Jun Zhang}

\institute{Max-Planck-Institut f\"{u}r Sonnensystemforschung,
D-37191, Katlenburg-Lindau, Germany \\
E-mail: zjun@ourstar.bao.ac.cn; solanki@mps.mpg.de; woch@mps.mpg.de
 \and
National Astronomical Observatories, Chinese Academy of
Sciences, Beijing 100012, China \\
E-mail: wjx@ourstar.bao.ac.cn}

\date{Received ??, 2007; accepted ??????}

\titlerunning{Moving Magnetic Features}

\authorrunning{Jun Zhang et al.}


\abstract{Using data recorded by the Michelson Doppler Imager
(MDI) instrument on the Solar and Heliospheric Observatory (SOHO),
we have traced 123 pairs of opposite magnetic polarity moving
magnetic features (MMFs) in three active regions NOAA ARs 8375,
0330 and 9575. At the time of observation, AR 8375 was young, AR
0330 mature, and AR 9575 decaying. Some properties of MMF pairs in
AR 8375 were studied by Zhang et al. (2003). The vertical
velocity, measured from MDI Dopplergrams for the three active
regions, indicates that the elements of MMF pairs with polarity
opposite to that of the sunspot support a downflow (Doppler
redshift) of around 50-100 m s$^{-1}$. The average Doppler shift
difference between negative and positive elements of an MMF pair
is about 150 m s$^{-1}$ in AR 8375, 100 m s$^{-1}$ in AR 0330, and
20 m s$^{-1}$ in AR 9575. These observational results are in
agreement with the model that MMF pairs are part of a U-loop
emanating from the sunspot's magnetic canopy. According to this
model the downflow is caused by the Evershed flow returning below
the solar surface. For AR 8375, the horizontal velocity of MMFs
ranges from 0.1 km s$^{-1}$ to 0.7 km s$^{-1}$, and on average,
the velocity of an MMF pair decreases significantly (from 0.6 km
s$^{-1}$ to 0.35 km s$^{-1}$) with increasing distance from the
MMF's birth place. In contrast, the decrease of the average
velocity is far less obvious from 0.5 km s$^{-1}$ to 0.4 km
s$^{-1}$ with increasing distance from the sunspot. This result
suggests that the change in MMF flow speed does not reflect the
radial structure of the moat flow, but rather is intrinsic to the
evolution of the MMF pairs. This result is also in agreement with
the U-loop model of MMF pairs. We also find that properties of MMF
pairs, most strikingly the lifetime, depend on the evolution
stages of the parent sunspot. The mean lifetimes of MMF pairs in
ARs 9575 and 0330 are 0.7 hours and 1.6 hours, respectively, which
is considerably shorter than the 4 hours lifetime previously found
for AR 8375.
}

\maketitle

\section{Introduction}

Moving magnetic features (MMFs) are small magnetic structures that
move away from a sunspot to the periphery of the surrounding moat
(Vrabec 1971; Harvey \& Harvey 1973; Muller \& Mena 1987;
Brickhouse \& LaBonta 1988; Lee 1992; Zhang et al. 2003; Hagenaar
\& Shine, 2005). These MMFs have been classified into three types
and their properties have been summarized by Shine \& Title
(2001); (see also Weiss et al. 2004, for a review). Type I MMFs
consist of bipolar pairs of magnetic elements. The bipolar pairs
move jointly outward across the moat at speeds of 0.5 $-$ 1 km s
$^{-1}$. They usually first appear just outside the sunspot along
a radial line extending from a dark penumbral filament, although
some MMFs originate inside penumbrae (Sainz Dalda \& Mart\'{i}nez
Pillet 2005; Zhang et al. 2007). Type II MMFs are single magnetic
elements with the same polarity as the sunspot, moving outward
across the moat at speeds similar to that of type I MMFs, while
type III MMFs are single magnetic elements with polarity opposite
to that of the sunspot, moving outward at significantly higher
speeds of 2$-$3 km s $^{-1}$.

Harvey \& Harvey (1973) proposed a model in which magnetic flux is
removed from the sunspot at the photospheric level. In this model
flux tubes form a sea serpent and MMFs are the intersections of
these flux tubes with the solar surface. An alternative
possibility was suggested by Wilson (1973, 1986; cf. Spruit et al.
1987). In his model, a thin magnetic flux tube is detached from
the main flux of the sunspot well below the surface. The detached
tube moves turbulently to the surface, developing twists and
kinks, which are seen as MMFs once it reaches the solar surface.
Also in this model a structure similar to a sea serpent can be
formed. Finally, Ryutova et al. (1998) have modelled MMF pairs as
${\Omega}$ loops emerging from below. They propose that these
loops are kinks of a horizontal flux tube lying below the surface.
They model the propagating kinks as a solitary wave.

Using Big Bear Solar Observatory, Yurchyshyn et al. (2001) studied
the longitudinal magnetic fields of 28 MMF pairs, associated with
two large sunspots. They find that MMFs are not randomly oriented.
The magnetic element having the same polarity as the sunspot is
located further from the sunspot than the opposite polarity
element. Furthermore, they find a correlation between the
orientation of the MMF bipoles and the twist of the sunspot
superpenumbra, as deduced from H${\alpha}$ images. Zhang et al.
(2003) confirmed the results of Yurchyshyn et al. (2001) for a
larger sample of MMF pairs, and deduced further systematics of
MMFs properties. MMFs tend to cluster at particular azimuths
around the parent sunspot and move approximately radially outward
from sunspots at an average speed of 0.45 km s$^{-1}$. Their
motion is deflected towards large concentrations of magnetic flux
of opposite polarity to that of the parent sunspot. Zhang et al.
argued that these and other observations are best reproduced by a
model in which MMFs are the intersections of U-loops, produced by
localized dips of the magnetic canopy surrounding sunspots, with
the solar surface.

Inside and around sunspots, many flows have been observed, which
may affect the velocity structure of MMFs (Solanki 2003). Such
flows are: 1. Evershed flow (Evershed, 1909), a predominantly
radial horizontal outflow seen in the penumbra (see Muller 1992;
Thomas 1994). Flow velocities of several km s$^{-1}$ (Bumba 1960;
Wiehr 1995; Schlichenmaier \& Schmidt 2000) and even supersonic
values (Borrero et al. 2005) have been reported in connection with
the Evershed flow. 2. moat flows, radial outflows around decaying
sunspots (Sheeley, 1969; 1972); 3. downflows near the outer
penumbral border and upflows near the inner penumbral border
(Westendorp Plaza et al. 1997; Hirzberger \& Kneer 2001; cf.
Tritschler et al. 2004).

In this paper we study mainly the horizontal and Doppler velocity
of MMF pairs around three sunspots in active regions NOAA 8375,
9575 and 0330. The velocity structure of MMFs provides additional
constraints that a successful model must satisfy. One aim of the
present paper is to test to what extent the model proposed by
Zhang et al. (2003) is able to reproduce these additional
observations. We also consider whether the properties of the MMFs
depend on the evolution stages of the sunspot.

\section{Observations and Analysis}

We combine magnetic field and Doppler velocity observations carried
out by the Michelson Doppler Imager, MDI (Scherrer et al. 1995) on
the Solar and Heliospheric Observatory (SOHO). MDI was employed in
the high-resolution mode (0.625 arcsec per CCD pixel and a 1 minute
cadence). Observations of three active regions, NOAA ARs 8375, 0330
and 9575, are analyzed. At the time of observations, the three
active regions were at three different evolutional stages. AR 8537
was young and still exhibited some flux emergence, AR 0330 was
regular and mature, and AR 9575 was also regular but decaying.
All three active regions were located near the central meridian
(N18W06 for AR 8375, N10E05 for AR 9575 and N09W01 for AR 0330).
Each active region
has a relatively large compact leading sunspot of positive polarity
and an extended negative polarity region.
Fig. 1 shows MDI continuum images (left) of ARs 8375 (top row),
AR 9575 (middle row) and AR 0330 (bottom row),
corresponding MDI longitudinal magnetograms (middle),
as well as corresponding MDI Dopplergrams (right).

In order to eliminate the Doppler signal caused by the 5-minute
oscillations, we averaged over five successive Dopplergrams. Thus,
Dopplergrams with a cadence of 5 minutes are analyzed. We
identified 42 MMF pairs in AR 8375 during 40 hours of observation
time (from 1998 November 23 18:53 UT to 25 10:51 UT), 55 MMF pairs
in AR 9575 during 11 hours (from 2001 August 16 19:38 UT to 17
06:48 UT) and 12 hours (from 2001 August 17 18:09 UT to 18 06:25
UT), respectively, and 26 MMF pairs in AR 0330 during 10 hours
(from 2003 April 9 14:05 UT to 10 00:48 UT).

The MMF pairs were identified by visually scanning successive
magnetograms. For a feature to be selected as an MMF, we required
it to appear in at least 10 magnetograms. We selected only
well-isolated MMF pairs. This may bias our selection towards
tighter pairs. The studied MMFs were located around the leading
sunspots.

Zhang et al. (2003) determined a set of parameters for each pair
of MMFs identified in two young active regions NOAA ARs 8375 and
9236. Here we extend their study by analysing a mature active
region (AR 0330), and a decaying one (AR 9575). This will help to
disentangle the properties of MMFs in active regions at different
stages of their evolution. In addition, we also consider new
physical parameters not considered by Zhang et al. Besides
analysing the location of first appearance and the lifetime of MMF
pairs, we concentrate on the velocity structure of MMF pairs.

\section{Location of first appearance and lifetime of MMFs in ARs
8375, 0330 and 9575}

For young active regions, the majority of MMF pairs first appears
at a distance of 1000 to 7000 km from the outer boundary of
sunspots. The mean distance at first appearance is 4500 km, with
standard deviation of 3800 km. The mean lifetime of MMFs is around
4 hours (Zhang et al. 2003), with standard deviation of 2.1 hours.

Fig. 2 shows the distribution of the distance to the sunspot
penumbral boundary at first appearance of 123 MMF pairs,
identified in the young active region (AR8375), the mature one (AR
0330), and the decaying one (AR 9575). Negative values mean that
the corresponding MMF pairs appear inside the penumbrae (i.e.
within the three closed dotted curves in the continuum images of
ARs 8375, 0330 and 9575 in Fig. 1). For mature and decaying active
regions, 12 out of the 81 MMF pairs were first seen within the
penumbral area, although we cannot rule out that some were missed
against the relatively strong penumbral signal. The mean distance
at first appearance of the 69 MMF pairs first observed outside the
penumbra is 3100 km with standard deviation of of 2300 km. This
distance is somewhat shorter than the mean distance of 4500 km
found in young active regions (see also Zhang et al. 2003).

By tracking MMF pairs in the three active regions from birth to
death, we have determined their lifetimes. Fig. 3 shows the
lifetime distribution of the 123 MMF pairs. For the MMFs in AR
9575, the lifetime ranges from 0.2 to 2.2 hours, with the peak of
the distribution close to 0.6 hour. The average lifetime is 0.7
hour with standard deviation of 0.3 hour. For the MMFs in AR 0330,
the lifetime ranges from 0.5 to 2.7 hours, and the mean lifetime
is 1.6 hours with standard deviation of 0.6 hour, which is longer
than that in the decaying active region, AR 9575. The lifetimes
for these two ARs are significantly shorter than the 4 hours found
for the young active region AR 8375 (e.g. Zhang et al. 2003).

\section{The velocity structure of MMF pairs}

\subsection{Doppler velocity}

An interesting parameter of MMF pairs is the Doppler shift inside MMF
elements. Fig. 4 presents examples of two individual MMF pairs. It shows MDI
line-of-sight magnetograms (left) and corresponding MDI Dopplergrams (right)
in the northern region of the sunspot of the young AR 8375 on Nov. 4, 1998.
The rectangular box on the magnetogram at 01:27 UT marks an MMF pair. The
time series of magnetograms and Dopplergrams in the lower frames exemplify
the horizontal motion of the MMF pairs within a 20 minute time interval as
well as the distribution and evolution of downflows (white patches),
respectively, upflows (black patches). The outward motion of the MMF is
clearly visible. Furthermore, the negative element of the MMF (i.e. the
polarity opposite to that of the parent sunspot) shows significant and
temporally stable downflow. Fig. 5 shows MDI magnetograms and corresponding
MDI Dopplergrams of the north-western region of the sunspot in the mature AR
0330. Again, the negative element of the MMF pair shows a downflow and the
magnitude of the flow is stable with time.

The left column of Fig. 6 displays the histogram of the Doppler
shift of the individual MMFs, identified from the three active
regions, with the thick line referring to MMF elements of positive
polarity (i.e. the same as the sunspot), the thin line to negative
elements. The figure also shows the difference of Doppler shift
between the negative and positive elements. For the young active
region NOAA AR 8375, the Doppler shift of elements of both
polarity ranges from $-$350 m s$^{-1}$ (blueshift) to 550 m
s$^{-1}$ (redshift). However, the distribution of the negative
polarity is shifted towards positive velocities. For the elements
of positive polarity, the peak is located near $-$50 m s$^{-1}$,
for the negative polarity element at $+$100 m s$^{-1}$. The
average difference of the Doppler shift between negative and
positive elements of an MMF pair is 150 m s$^{-1}$, as shown in
the top-right frame of Fig. 6, and the standard deviation of the
Doppler shift of an MMF element is about 55 m  s$^{-1}$. This
implies that plasma in the negative element moves downward
relative to that in the positive element, and the difference is
significant. For the mature active region AR 0330, the Doppler
shift of MMFs ranges from $-$300 m s$^{-1}$ to 550 m s$^{-1}$. The
peak is located near 0 m s$^{-1}$ for positive polarity elements,
and 100 m s$^{-1}$ for negative polarity elements. The difference
is 100 m s$^{-1}$, with standard deviation of roughly 36 m
s$^{-1}$. Again, plasma in the element closer to the sunspot moves
downward relative to that in the more remote element. The Doppler
shift of MMFs in the decaying active region (AR 9575) ranges from
$-$ 170 m s$^{-1}$ to 280 m s$^{-1}$, and the peak is located at
20 m s$^{-1}$ and 40 m s$^{-1}$ for positive and negative polarity
elements, respectively. There is a smaller difference (20 m
s$^{-1}$) of the Doppler shifts between the negative and positive
elements, compared to those of the young and mature active
regions. This difference is below the 1${\sigma}$ level (24 m
s$^{-1}$) and is not significant.

We have also studied the evolution with time of the difference in
Doppler shifts between positive and negative elements. Generally,
for a given MMF pair the difference of the Doppler shifts between
the negative and positive elements is remarkably stable as a
function of time. E.g. for AR 8375, the difference at first
appearance is 157 m s$^{-1}$, and at last detection, 151 m
s$^{-1}$

The zero level for the line-of-sight velocity in Fig. 6 was set by taking the
average Doppler shift of a quiet-Sun region at the same longitude and setting
this to zero. This implies that the zero level corresponds to a small
redshift (typically 200-300 m s$^{-1}$ for photospheric lines of neutral
metals), since the granular blue shift of the Ni I line hasn't been removed.
This means that the downflow in the negative polarity MMFs is larger than
indicated by this figure.

\subsection{Evolution of the horizontal velocity}

In general, MMF pairs move roughly radially away from the center of the
parent sunspot. Zhang et al. (2003) reported that the average horizontal
velocity is 0.45 km s$^{-1}$ in the young active region AR 8375. Here we
analyse the evolution of the horizontal velocity with distance from the
penumbral boundary and from their birth place. Due to the short lifetime of
MMFs in ARs 9575 and 0330, these MMFs move only a shorter distance before
disappearing, so that it is difficult to measure changes in their horizontal
speed. We therefore only analysed the velocity of the 42 MMF pairs in AR
8375. The velocity is determined by measuring the distance travelled by each
MMF element in an interval of about 1 hour. At each instance we also recorded
the nearest distance to the sunspot penumbral boundary. Fig. 7 shows the
relationship between horizontal velocity and distance to the sunspot penumbra
of AR 8375, separately for positive (top) and negative (bottom) elements. The
horizontal velocity decreases from an average value of 0.5 km s$^{-1}$ to 0.4
km s$^{-1}$, as the distance to the sunspot penumbra increases from 2000 km
to 12000 km. However, the scatter is large and the trend not entirely clear.
This is reflected by the low absolute values of the correlation coefficient
(given in the figure).

Fig. 8 displays the horizontal velocity of MMF elements versus relative
distance from MMF birth place. The velocity decreases from ${\sim}$ 0.6 km
s$^{-1}$ to 0.35 km s$^{-1}$, with a strongly increased anti-correlation
between velocity and relative distance from MMF birth place. This means that
usually MMF elements have a higher radial velocity at their first appearance,
but slow down while moving from the sunspot to the periphery of the
surrounding moat.

The main source of error of velocity determination of MMF elements is due to
uncertainty in the MMF position. A position error of one pixel introduces an
error of the velocity of (one pixel)/(time interval). As the size of one
pixel is about 0.6 arcsec and the time interval is about 1 hour, the error in
velocity becomes 0.1 km s$^{-1}$, which may explain a part of the scatter in
Figs. 7 and 8.

Note that the MMF velocity found by us, at least during the later stages of
an MMF pair's life, are similar to those of intranetwork magnetic elements
(Zhang et al. 1998), which are thought to be dragged along by the
supergranular flow. The initially higher speed of the MMFs may have two
causes: Either the moat flow is more vigorous closer to the sunspot, or the
MMFs are initially carried outwards partly by the momentum of the Evershed
flow, which has speeds of 1$-$2 km s$^{-1}$ in the canopy of a sunspot
(Solanki et al. 1994) before aerodynamic drag slows them down to the ambient
speed of the moat flow. In the first case we expect the speed of the MMFs to
decrease mainly as a function of distance to the sunspot. In the latter case
we expect the MMF speed to decrease as a function of distance from the point
at which the MMF was formed. The results of our study clearly favour the
second hypothesis, which is consistent with the U-loop model of MMF pairs
proposed by Zhang et al. (2003).

\section{Discussions and Conclusions}

Zhang et al. (2003) reported that MMF bipoles are not randomly oriented (see
also Yurchyshyn et al. 2001), but rather that the member of an MMF pair
further from the sunspot has the polarity of the parent sunspot in 85\% of
the cases, the orientations of MMF pairs are further associated with the
twist of the sunspot superpenumbra. This supports the picture that MMF pairs
are parts of U-loops `hanging' below a sunspot's superpenumbral canopy (see
Fig. 9 of Zhang et al. 2003). Also, the separation between the two polarities
of an MMF pair does not change significantly with time in contrast to what
one would expect for an emerging ${\Omega}$ loop. Finally, more MMF pairs are
seen in the direction of the opposite polarity pore/sunspot in AR 8375, a
direction in which the canopy is expected to lie particularly low.

In the model of Zhang et al. (2003) the Evershed flow plays an important role
in the formation of MMF pairs. At the edge of the penumbra this supporting
force disappears and a sufficiently dense and massive packet of Evershed gas
cannot be supported by the flux-tube field any more. This gas then sinks,
taking the magnetic field with it. In this way a U-loop is created near the
penumbral edge, so MMF pairs first appear just outside the penumbrae (see
Fig. 2). Alternatively, the flux tube can sink already within the penumbra,
if it gets deformed downward. Such a deformation leads to a downflow, which
makes this part of the tube more dense, making it bend down more, and so on,
leading to an instability (see Schlichenmaier 2002), so that basically the
same mechanism can lead to the production of MMFs outside the penumbra, or
inside it (Sainz Dalda \& Mart\'{i}nez Pillet 2005; Zhang et al. 2007). We
expect the footpoint with opposite polarity to that of the sunspot to show
signs of this downflowing material. Since the gas density increases rapidly
with depth we do not expect a corresponding upflow in the other footpoint.
Figs. 4 and 5 show two examples of MMF pairs in which the negative element
(close to the sunspot with positive polarity) has a downflow. Fig. 6 confirms
that this is a general phenomenon. The majority of the negative elements show
a significant downflow compared to the positive polarity.

This downflow may be related to the isolated downflow seen just outside a
sunspot by B\"{o}rner and Kneer (1992) and may partly explain why the
Evershed velocity in the superpenumbra does not increase with distance from
the sunspot, although the increasing canopy height implies that the mass flux
must decrease (Solanki et al. 1994; 1999): a part of the mass drains down
into MMFs. The rather low MMF downflow velocities suggest that the field in
the MMF pairs is heavily inclined to the vertical. This property is shared
with the field lines found to submerge at the edge of the penumbra by
Westendorp Plaza et al. (1997) and Mathew et al. (2002). By analogy to the
MMF pairs we therefore expect such field lines also to eventually rejoin the
magnetic canopy. This picture is supported by the simulations of
Schlichenmaier (2002). Recently, Cabrera Solana et al. (2006) provide strong
evidence that at least some MMFs are the continuation of the penumbral
Evershed flow into the moat, this gives further support to the earlier
results of Zhang et al. (2003) about the magnetic connection between MMFs and
the penumbra. Note that in other models of MMFs, that do not involve U-loops
from the canopy, a downflow predominantly in the MMF pair member closer to
the spot is less easy to accommodate in a natural way.

The observation that the MMFs initially move faster early in their life
before slowing down (see Fig. 8), suggests that although the moat flow may be
the prime driver of older MMFs, other mechanisms, e.g. the Evershed flow,
help drive younger MMFs. Changes in the moat flow cannot be the main cause of
this deceleration, since the MMFs' velocity correlates poorly with the
distance from the sunspot (Fig. 7). These results further confirm our MMF
model (Zhang et al. 2003) that MMF pairs, i.e. type I MMFs are formed when
the field lines in a small part of the magnetic canopy dip down to produce a
U-loop.

Besides looking for further evidence to test models of MMFs we have also
checked if the properties of MMFs change between young and old sunspots. The
second main result we find is that the lifetime of MMFs around young sunspots
is quite a bit longer than around older ones.

Also, the downflows in MMFs are more pronounced around younger
sunspots. Specifically, for the young AR studied here nearly all
MMF elements with polarity opposite to the sunspot show a
significant redshift compared to the other elements. Of course,
many more sunspots at different stages of their development need
to be studied before we can be certain that MMF properties do
depend on evolution stages of sunspots.

\begin{acknowledgements}
The authors are indebted to the SOHO/MDI team for providing the
employed data. SOHO is a mission of international cooperation
between ESA and NASA. JZ was supported by the National Natural
Science Foundations of China (G10573025 and 40674081), the CAS
Project KJCX2-YW-T04, the National Basic Research Program of China
under grant G2006CB806303, and the cooperation agreement between
the Chinese Academy of Sciences and the Max-Planck Society.
\end{acknowledgements}

\clearpage

\begin{figure*}
\caption{{\it
Left:} MDI continuum images of NOAA AR 8375 (top row), NOAA AR
9575 (middle row) and NOAA AR 0330 (bottom row); {\it middle:}
corresponding MDI longitudinal magnetograms; {\it right:}
corresponding MDI Dopplergrams. The dotted curves in the continuum
images outline the boundary of the penumbrae. }
\end{figure*}

%

\begin{figure*}
\caption{Histogram of the distance of first appearance of MMF
pairs from the sunspot boundary of the three active regions ARs
9575, 0330 and 8375. }
\end{figure*}

\begin{figure*}
\caption{Histogram of MMF lifetimes in the three active regions
ARs 9575, 0330 and 8375. }
\end{figure*}

\begin{figure*}
\caption{{\it
Left:} SOHO/MDI longitudinal magnetograms showing the northern
part of the sunspot of AR NOAA 8375 on Nov. 4, 1998. A box at
01:27 UT outlines the field-of view of the following magnetograms.
The arrows denote an MMF pair; {\it right:} corresponding MDI
Dopplergrams. White patches show downflow, and black patches,
upflows. The arrows point to the location of the MMF pair. }
\end{figure*}

\begin{figure*}
\caption{Similar to Fig. 4 but for AR 0330. {\it Left:} MDI
longitudinal magnetograms showing the north-western part of the
sunspot; {\it right:} corresponding MDI Dopplergrams. }
\end{figure*}

\begin{figure*}
\caption{{\it
Left column:} histogram showing the Doppler shift of the
individual MMFs in the three active regions; {\it right column:}
histogram showing the difference of the Doppler shift between the
negative and positive MMF elements in the three active region. }
\end{figure*}

\begin{figure*}
\caption{The relationship between horizontal
velocity and distance to the sunspot penumbra of AR NOAA 8375.
{\it Upper panel:} horizontal velocity of the positive elements of MMF
pairs; {\it lower panel:} horizontal velocity of the negative elements.
In each panel, the solid line represents a linear regression, with the
1${\sigma}$ uncertainty
of the velocity outlined by two dotted lines, ${\alpha}$ represents the
correlation coefficient.
}
\end{figure*}

\begin{figure*}
\caption{The
horizontal velocity of MMF elements is plotted versus relative
distance from MMF birth place. Otherwise the figure is similar to
Fig. 7. }
\end{figure*}



\begin{thebibliography}{}
\bibitem{} B\"{o}rner, P., \& Kneer, F. 1992, A\&A, 259, 307
\bibitem{} Borrero, J. M., Lagg, A., Solanki, S. K., \& Collados, M.
2005, A\&A, 436, 333
\bibitem{} Brickhouse, N. S.,
\& LaBonta, B. J. 1988, Sol. Phys., 115, 43
\bibitem{} Bumba, V. 1960, Izv. Crim. astrophys. Obs., 23, 253
\bibitem{} Cabrera Solana, D., Bellot Rubio, L. R., Beck, C., \& del
Toro Iniesta, J. C. 2006, ApJ, 649, L41
\bibitem{} Evershed, J. 1909, MNRAS, 69, 454
\bibitem{} Hagenaar, H. J., \& Shine, R. A. 2005, ApJ, 635, 659
\bibitem{} Harvey, K., \& Harvey, J. 1973,
Sol. Phys., 28, 61
\bibitem{} Hirzberger, J., \& Kneer, F. 2001, A\&A, 378, 1078
\bibitem{} Lee, J. W. 1992, Sol. Phys., 139, 267
\bibitem{} Mathew, S. K., Solanki, S. K., Lagg, A., et al. 2002,
in Poster Proc. 1st Potsdam Thinkship on Sunspot \& Starspots,
K. Strassmeier (Ed.), AIP, 117
\bibitem{} Muller, R., \& Mena, B. 1987,
Sol. Phys., 112, 295
\bibitem{} Muller, R. 1992, in NATO ASI Proc. 375: Sunspots. Theory
and Observations, ed. J. H. Thomas, \& N. O. Weiss, Kluwer, Dordrecht, 175
\bibitem{} Ryutova, M., Shine, R., Title, A.,
\& Sakai, J. I. 1998, ApJ, 492, 402
\bibitem{} Sainz Dalda, A., \& Mart\'{\i}nez Pillet, V.
2005, \apj, 632, 1176
\bibitem{} Scherrer, P. H., Bogart, R. S.,
Bush, R. I., et al. 1995, Sol. Phys., 162, 129
\bibitem{} Schlichenmaier, R. 2002, Astron. Nachr., 323, 342
\bibitem{} Schlichenmaier, R., \& Schmidt, W. 2000, A\&A, 358, 1122
\bibitem{} Sheeley, N. R. Jr. 1969, Sol. Phys., 9, 347
\bibitem{} Sheeley, N. R. Jr. 1972, Sol. Phys., 25, 98
\bibitem{} Shine, R. \& Title, A. 2001, Encyclopedia of A\&A,
Vol. 4, 3209
\bibitem{} Solanki, S. K. 2003, A\&A Rev, 11, 153
\bibitem{} Solanki, S. K., Montavon, C. A. P., \& Livingston, W. 1994,
A\&A, 283, 221
\bibitem{} Solanki, S. K., Finsterle, W., R\"{u}edi, I.,
\& Livingston, W. 1999, A\&A, 347, L27
\bibitem{} Spruit, H. C.,
Title, A. M., \& van Ballegooijen, A. A. 1987, Sol. Phys., 110, 115
\bibitem{} Thomas, J. H. 1994, in Solar Surface Magnetism, ed. R. J.
Rutten, \& C. J. Schrijver, Kluwer, Dordrecht, 219
\bibitem{} Tritschler, A., Schlichenmaier, R., Bellot Rubio, L. R. et
al. 2004, A\&A, 415, 717
\bibitem{} Vrabec, D. 1971, in R. Howard (ed.) Solar
Magnetic Fields, IAU Symp. 43, Reidel, Dordrecht, 329
\bibitem{} Weiss. N. O., Thomas, J. H., Brummell, N. H., \& Tobial, S.
M. 2004, ApJ, 600, 1073
\bibitem{} Westendorp Plaza, C., del Toro Iniesta, J. C.,
Ruiz Cobo, B., et al., 1997, 1st Advances in Solar Physics
Euroconference. Advances in Physics of Sunspots, ASP Conf. Ser.
Vol. 118., Eds.: B. Schmieder, J.C. del Toro Iniesta,
\& M. Vazquez, p. 202
\bibitem{} Wiehr, E. 1995, A\&A, 298, L17
\bibitem{} Wilson, P. R. 1973, Sol. Phys., 32, 435
\bibitem{} Wilson, P. R. 1986, Sol. Phys., 106, 1
\bibitem{} Yurchyshyn, V. B.,
Wang, H., \& Goode, P. R. 2001, ApJ, 550, 470
\bibitem{} Zhang, J., Wang, J., Wang, H., \&
Zirin, H. 1998, A\&A, 335, 341
\bibitem{} Zhang, J., Solanki, S. K., \& Wang, J. 2003, A\&A, 399, 755
\bibitem{} Zhang, J., Solanki, S. K., \& Woch, J. 2007, A\&A Letters, submitted
\end{thebibliography}
\end{document}